\documentclass[pre, twocolumn, amsmath, amssymb, superscriptaddress]{revtex4}   
\usepackage{multirow}
\usepackage{graphicx}
\usepackage{dcolumn}
\usepackage{bm}
\usepackage{braket}
\usepackage{xcolor}

\usepackage{amsmath}

\newcommand{\beq}{\begin{equation}}
\newcommand{\eeq}{\end{equation}}

\usepackage{xr}
\usepackage{caption}

\usepackage{setspace}
\usepackage[a4paper,margin=25mm]{geometry}
\begin{document}

\title{How genome redundancy can promote evolutionary innovation}

\author{Domenico Caudo}
\affiliation{Department of Physics, Sapienza University of Rome, Rome, Italy}
\affiliation{Center for Life Nano \& Neuro Science, Italian Institute of Technology, Rome, Italy}

\author{Mattia Miotto \footnote{\label{corr} For correspondence write to: greta.grassmann@uniroma1.it and mattia.miotto@uniroma1.it}}
\affiliation{Department of Physics, Sapienza University of Rome, Rome, Italy}
\affiliation{Center for Life Nano \& Neuro Science, Italian Institute of Technology, Rome, Italy}

\author{Giancarlo Ruocco}
\affiliation{Center for Life Nano \& Neuro Science, Italian Institute of Technology, Rome, Italy}
\affiliation{Department of Physics, Sapienza University of Rome, Rome, Italy}

\author{Greta Grassmann \ref{corr}}
\affiliation{Department of Physics, Sapienza University of Rome, Rome, Italy}
\affiliation{Center for Life Nano \& Neuro Science, Italian Institute of Technology, Rome, Italy}

\begin{abstract}
Polyploidy is defined as the existence of more than two complete sets of homologous chromosomes. 
Despite it being a widespread phenomenon across the tree of life, its role as either an evolutionary innovation or a dead end is still debated. 
Here, we investigate how under varying selective pressures the degree of ploidy interacts with two key biological factors: the mode of inheritance and the genotype-phenotype mapping.
Through a minimal evolutionary model we find that polyploidy is especially advantageous during abrupt environmental changes, confirming that polyploidization is often associated with ecological upheavals. We observe that stochastic inheritance combined with a nonlinear (maximum-based) genotype-phenotype mapping maximizes both phenotypic exploitation and landscape exploration across all environments. 
By contrast, structured inheritance with an additive phenotype mapping systematically underperforms, yet displays a pronounced optimum at low-to-intermediate ploidy level that 
mirrors the distribution observed in natural plant and bacterial populations.
When individuals are free to carry different chromosome numbers, selection drives the population toward values that reflect an interplay between exploitation, exploration, and convergence speed rather than any single evolutionary objective. The relative weight of these three factors depends on the fitness landscape, providing a unifying framework for understanding when and why polyploidy is favored by natural selection.
\end{abstract}

\flushbottom
\maketitle

\section*{Introduction}
Starting from the Last Universal Common Ancestor (LUCA), estimated to have lived around 4 billion years ago \cite{moody2024nature}, the tree of life has diversified progressively as organisms adapted to a wide range of environmental conditions through mutation, selection, and speciation \cite{hedges2015tree}. This process is shaped by both intrinsic factors, such as genetic drift and mutation \cite{grassmann2025insights,miotto2023differences}, and extrinsic forces imposed by the surrounding environment \cite{nei2007new,starr2025disentangling}.\\
Among the many factors shaping evolutionary dynamics, one that is still not completely understood is polyploidy \cite{van2017evolutionary,van2021polyploidy}.
Polyploid organisms possess more than two sets of homologous chromosomes as a result of whole-genome duplication (WGD), which can arise from abnormal cell division, interspecific hybridization, or errors during meiosis \cite{soltis2016polyploidy,mable2011genome,otto2000polyploid}.
The role of polyploidy has long been debated: is polyploidy a creative force that promotes speciation and adaptation, or does it represent an evolutionary dead end?
Comparative genomic analyses have suggested that polyploid lineages may experience higher extinction rates, indicating that many polyploidization events fail to give rise to persistent lineages \cite{arrigo2012rarely,mayrose2015methods,vanneste2014tangled}.
However, over the past decades, evidence for hundreds of WGD events during the last 500 million years has been uncovered, with many more likely to have occurred \cite{van2017evolutionary}. Moreover, at least one round of polyploidy has been identified in the ancestry of most plant lineages ($15$-$30\%$ of all angiosperm speciation events are associated with ploidy increases \cite{wood2009frequency}), and numerous extant examples exist in insects, fishes, amphibians, and reptiles \cite{mable2011genome,otto2000polyploid}.
It remains an open question whether polyploid lineages persist largely by chance, appearing randomly across the tree of life, or whether WGD events are preferentially associated with periods of ecological upheaval, environmental change, or mass extinction, contexts in which the increased genetic redundancy of polyploids may confer greater tolerance to a wider range of environmental conditions \cite{te2012more,hahn2012increased,ramsey2011polyploidy,madlung2013polyploidy}.\\
The evolutionary consequences of this increased redundancy are themselves debated. On one hand, polyploidy has been proposed to buffer the phenotypic effects of mutations: the presence of multiple gene copies may dilute the impact of deleterious or beneficial mutations \cite{koch1984evolution}. In light of Fisher’s fundamental theorem of natural selection, which relates the rate of evolution to the genetic variance in fitness \cite{fisher1999genetical}, this would suggest a reduced evolutionary rate in polyploids.
On the other hand, gene duplication provides raw material for evolutionary innovation: redundant gene copies can accumulate mutations without compromising essential functions, thereby facilitating the emergence of novel traits in a process known as neofunctionalization, as originally proposed by Ohno \cite{ohno2013evolution}.\\

Beyond the question of genetic redundancy, the evolutionary dynamics of polyploids depend critically on two additional biological ingredients that have received less theoretical attention: the genotype-phenotype mapping and the mode of inheritance \cite{hatakeyama2024evolutionary}.
In polyploid organisms, inheritance can follow different routes, ranging from disomic (pairwise chromosome segregation) to polysomic inheritance (random pairing and segregation among multiple homologous chromosomes) \cite{scott2023inference,le2010making,stift2008segregation}.
For example, newly formed polyploids often exhibit polysomic inheritance \cite{le2010making} and polyploid simple organisms such as bacteria replicate chromosomes sequentially rather than synchronously, leaving more room for stochastic (polysomic-like) segregation of chromosome copies \cite{ohbayashi2019coordination}. On the other hand, a more structured inheritance is usually observed in eukaryotes \cite{yanagida2005basic}.
The mapping between genotype and phenotype is equally non-trivial. Phenotypic traits are often described by a linear additive model, where the phenotype is the mean contribution of multiple loci \cite{hill2008data,hatakeyama2024evolutionary}.
However, dominance relationships and gene regulation thresholds can result in phenotypes better described by the maximum allelic contribution \cite{lynch2007origins,wagner2011pleiotropic,billiard2021integrative}.\\

To investigate how the natural variability of these two biological factors can influence the evolution of populations with different ploidy degrees, we compare two extreme inheritance modes coupled with two genotype-phenotype mapping strategies.
We then study how organisms adopting these four models behave across various types of environmental pressure by introducing several fitness landscapes and following the evolution of populations under each of them. 
We then extend the model to populations with heterogeneous number of chromosome copies and show that the evolutionary success of a given ploidy level reflects a context-dependent tradeoff between phenotypic exploitation (proximity to the fitness optimum), exploration (phenotypic variance at steady state), and convergence speed (generations to reach stationarity).


\section*{Results}

\subsection*{Model}
To describe the evolution of a population of fixed size $M$ of individuals, each carrying $N$ chromosome copies, we introduce the one-dimensional model shown in Figure \ref{fig1}a (see Methods for details). 
We consider a simplified genetic architecture in which the genotype is determined by the expression of a single locus present on the $N$ homologous copies. The corresponding gene product is transcribed and translated from all copies at equal rates, as observed experimentally in polyploid systems \cite{ohbayashi2019coordination,zheng2017cyanobacteria}. The activity of each copy is described by a scalar variable $x_{ij}(t) \in [0,1]$, which represents the quantitative trait (such as the intensity of scale coloration) of the $j$-th chromosome in the $i$-th individual at generation $t$. 
The genotype of an individual is defined as the set $\{x_{ij}(t)\}_{j=1}^N$, while the phenotype $y_i(t)$ is obtained as a function of this set through a mapping $y_i(t) = \Phi(\{x_{ij}(t)\})$.
To probe how different genotype-phenotype mappings influence evolutionary outcomes, we explore two alternative phenotype definitions: one based on the average (\textit{mean} mode) and one based on the maximum (\textit{max} mode) of all gene copies.
 The former is defined as
\begin{equation}
    y_i(t) = \sum_{j=1}^N\frac{x_{ij}(t)}{N},
\end{equation}
the latter as
\begin{equation}
    y_i(t) = max_j\{x_{ij}(t)\}.
\end{equation}
The fitness of individual $i$ is then given by a function of its phenotype $f(y_i(t))$, which reflects how successful (high fitness) or unsuccessful (low fitness) the organism's characteristics are in its environment, since it determines its reproductive success.
Indeed, in our model evolution proceeds through a selection–reproduction process reminiscent of a Wright–Fisher scheme with selection \cite{fisher1999genetical}: for each individual, a random number $u \sim \mathcal{U}[0,1]$ is drawn, and the individual is selected as a parent if $u < f(y_i(t))$.\\
Throughout the study we investigate three classes of fitness landscapes, representing increasing degrees of environmental complexity: the \textit{neutral fitness} provides a baseline with no selective optimum; the \textit{smooth fitness} introduces a single, well-defined global optimum modeling stabilizing selection; and the \textit{rough fitness} introduces multiple competing optima, probing the ability of populations to navigate complex adaptive landscapes and escape fitness valleys. A more detailed description is given in the following:
\begin{enumerate}

\item \textit{Neutral fitness}: $f(y_i(t)) = y_i(t)$. It models a neutral environment in which there are no additional constraints except the fixed number of individuals, so that individuals with higher phenotypic values have a proportionally higher probability of reproduction (e.g. more intensely colored fish are preferentially selected for reproduction).

\item \textit{Smooth fitness}: $f(y_i) = A\left(1 + \sin\left(B y_i - C\right)\right)$, with $A=0.5$, $B=2\pi$, and $C=0.5\pi$. This function defines a smooth, unimodal landscape maximized on a single global optimum at $y_i = 0.5$, and corresponds to a minimal model of stabilizing selection commonly used in theoretical evolutionary biology \cite{hatakeyama2024evolutionary}. Biologically, it can be interpreted as the result of competing selective pressures: for instance, an intermediate scale pigmentation level may maintain attractiveness to mates while minimizing predation risk \cite{matin1991molecular}.

\item \textit{Rough fitness}: given by a sum of gaussian functions $f(y_i)=A\mathcal{N}(\mu_A,\sigma_A)+B\mathcal{N}(\mu_B,\sigma_B)$. This is a multimodal fitness landscape characterized by both global and local maxima (depending on the parameters choice), which is relevant to real evolutionary transitions: fitness valley crossing has often been discussed in theoretical genomics \cite{komarova2012accelerated,weinreich2005rapid,weissman2009rate,hatakeyama2024evolutionary} to study evolutionary innovation, which is considered to have appeared when populations are observed to drastically change a character expression.\\

\end{enumerate}

Offspring are generated from selected parents according to the chosen inheritance scheme, and the procedure is iterated until $M$ offspring are produced.
To account for inheritance variability, we consider two extreme modes introduced by T. S. Hatakeyama \textit{et al.} \cite{hatakeyama2024evolutionary}: the \textit{set} mode, in which all gene copy from the mother are duplicated individually and inherited by daughters in the same manner and the \textit{random} mode, in which gene copies are sampled independently until the target number of genes is reached (capturing the effect of polysomic segregation).\\
By combining the two definitions of genotype-phenotype mapping and the two inheritance modes we build the \textit{mean-set}, \textit{mean-random}, \textit{max-set}, and \textit{max-random} models.\\ 
Mutations are introduced at each generation as follows. For each individual, the number $k$ of mutated gene copies is drawn from a Poisson distribution $k \sim \mathrm{Poisson}(\lambda N)$. A subset of $k$ chromosomes is randomly selected and their gene values are reassigned by sampling from a Beta distribution (see Methods). This mutation scheme implements a diffusion process in genotype space with occasional large jumps, commonly used in standard models of quantitative trait evolution under stabilizing selection \cite{lande1976natural}.
We follow the evolution of the population until it converges to a stationary state, defined as the generation $t^*$ when the absolute difference between the mean phenotype at generation $t^*$ and the average mean phenotype over the preceding 500 generations is smaller than $10^{-3}$.\\
Results are averaged over 100 independent repetitions for each model and each $N \in [1, 100]$.

\subsection*{Stochastic inheritance and nonlinear phenotype mapping maximize exploitation and exploration in neutral environments}

\begin{figure*}[ht]
\centering
\includegraphics[width=\linewidth]{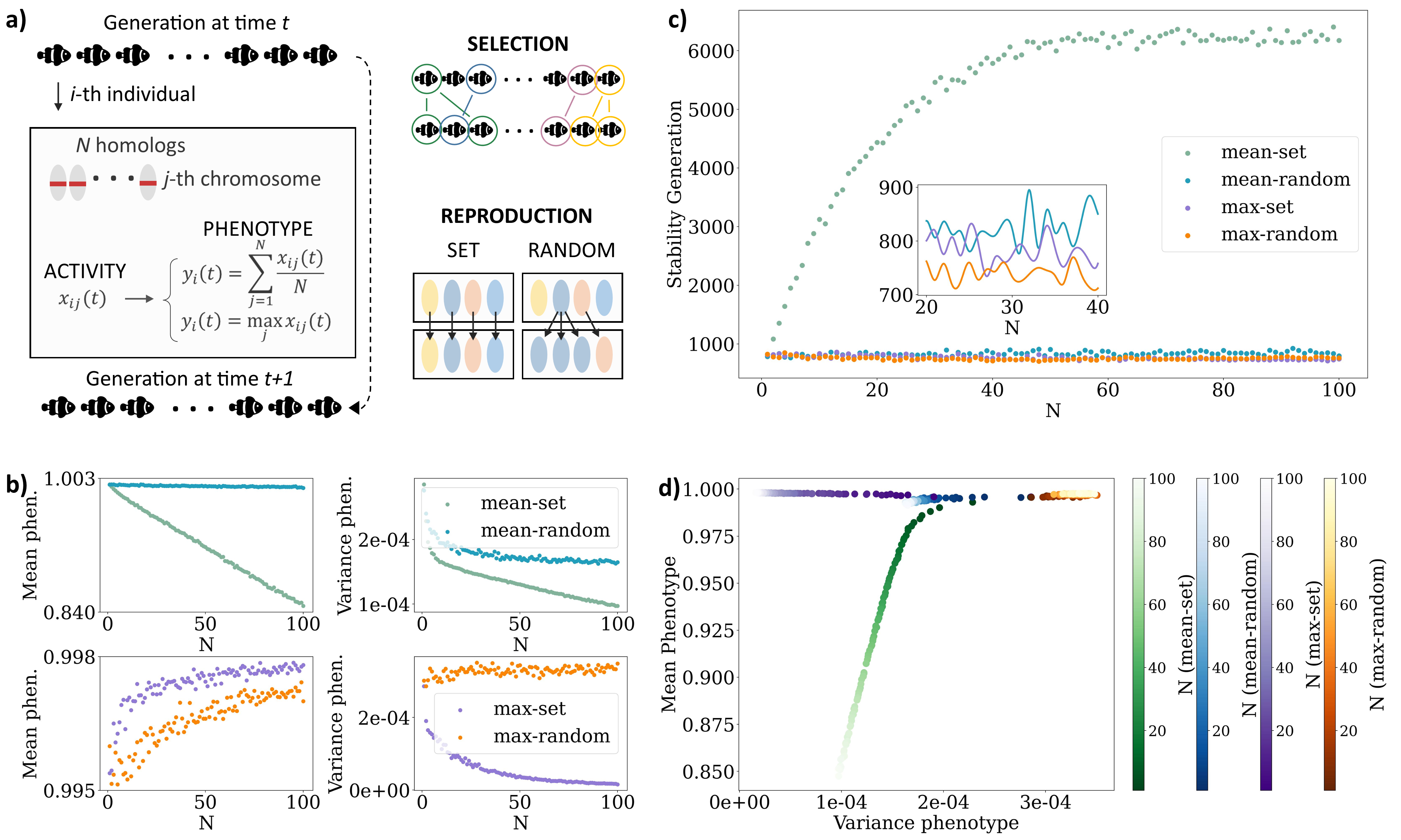}
\caption{\textbf{One-dimensional evolutionary model under four combinations of inheritance mode and phenotype mapping.}
\textbf{(a)} Schematic representation of the model, illustrating the selection and reproduction processes that drive the transition from one generation to the next. Each individual carries $N$ gene copies whose values determine the phenotype through either a \textit{mean} or \textit{max} mapping; reproduction follows either a \textit{set} or \textit{random} inheritance scheme.
\textbf{(b)} Mean phenotype (left column) and variance of the phenotype (right column) at steady state as a function of the ploidy level $N$. Top row: \textit{mean-set} (green) vs. \textit{mean-random} (blue). Bottom row: \textit{max-set} (violet) vs. \textit{max-random} (orange).
\textbf{(c)} Number of generations required to reach stability as a function of $N$ for each of the four models. In the inset, a zoomed portion of the splined interpolation of the stability generation values of the \textit{mean-random} and \textit{max} models.
\textbf{(d)} Mean vs. variance of the phenotype at steady state for each $N$ and each model, using the same color code.}
\label{fig1}
\end{figure*}

We first consider a minimal neutral environment modeled through the \textit{neutral fitness}.\\
In both \textit{mean-set} and \textit{mean-random} models, for small $N$ the population reaches a stationary state with mean phenotype close to one.
However, for increasing $N$, the two inheritance modes diverge significantly (Figure \ref{fig1}b): in the \textit{mean-random} model the mean phenotype remains approximately constant, whereas in the \textit{mean-set} model it decreases with $N$. This behavior follows directly from the law of large numbers: averaging over a larger number of independent copies makes it almost impossible to have fitness close to the maximum value. Indeed, this would imply for all copies of the gene to have a value close to one, which is however impossible due to the presence of mutations.
As a consequence, the system becomes less sensitive to beneficial fluctuations.
Consistently, the variance of the phenotype decreases with $N$ in both \textit{mean} models, but is systematically lower in the \textit{mean-set} case. Since variability is a key driver of adaptive potential \cite{fisher1999genetical,caudo2026robust}, we can interpret this behavior as a reduction in evolvability \cite{hansen2008measuring}.
By contrast, in the \textit{max} models both \textit{set} and \textit{random} inheritance lead to efficient phenotypic exploitation (Figure \ref{fig1}b). Increasing $N$ improves the mean phenotype in both cases, as the probability of sampling at least one high-value copy increases with the number of homologues.
The \textit{max-random} model additionally shows a variance largely independent of $N$ and approximately three times as large as in the other three models, indicating sustained exploration of the phenotypic landscape.\\
The differences between models are also reflected in convergence speed, as shown in Figure \ref{fig1}c. The \textit{mean-set} model converges significantly more slowly (up to six times slower for large $N$) indicating inefficient exploration of the fitness landscape.\\

Overall, these results suggest that, in the absence of environmental constraints, evolutionary success is maximized by a combination of stochastic inheritance and nonlinear phenotype mapping. 
As summarized in Figure \ref{fig1}d, the \textit{mean-set} model clearly deviates from the others, showing reduced exploitation and exploration as $N$ increases. The remaining three models converge to similar mean phenotypes, but the \textit{max-random} model provides both high fitness and sustained variability, two key ingredients for long-term success \cite{draghi2008evolution,de2019exploration}.

\subsection*{An optimal ploidy level emerges under stabilizing selection in the \textit{mean-set} model}

\begin{figure*}[ht]
\centering
\includegraphics[width=\linewidth]{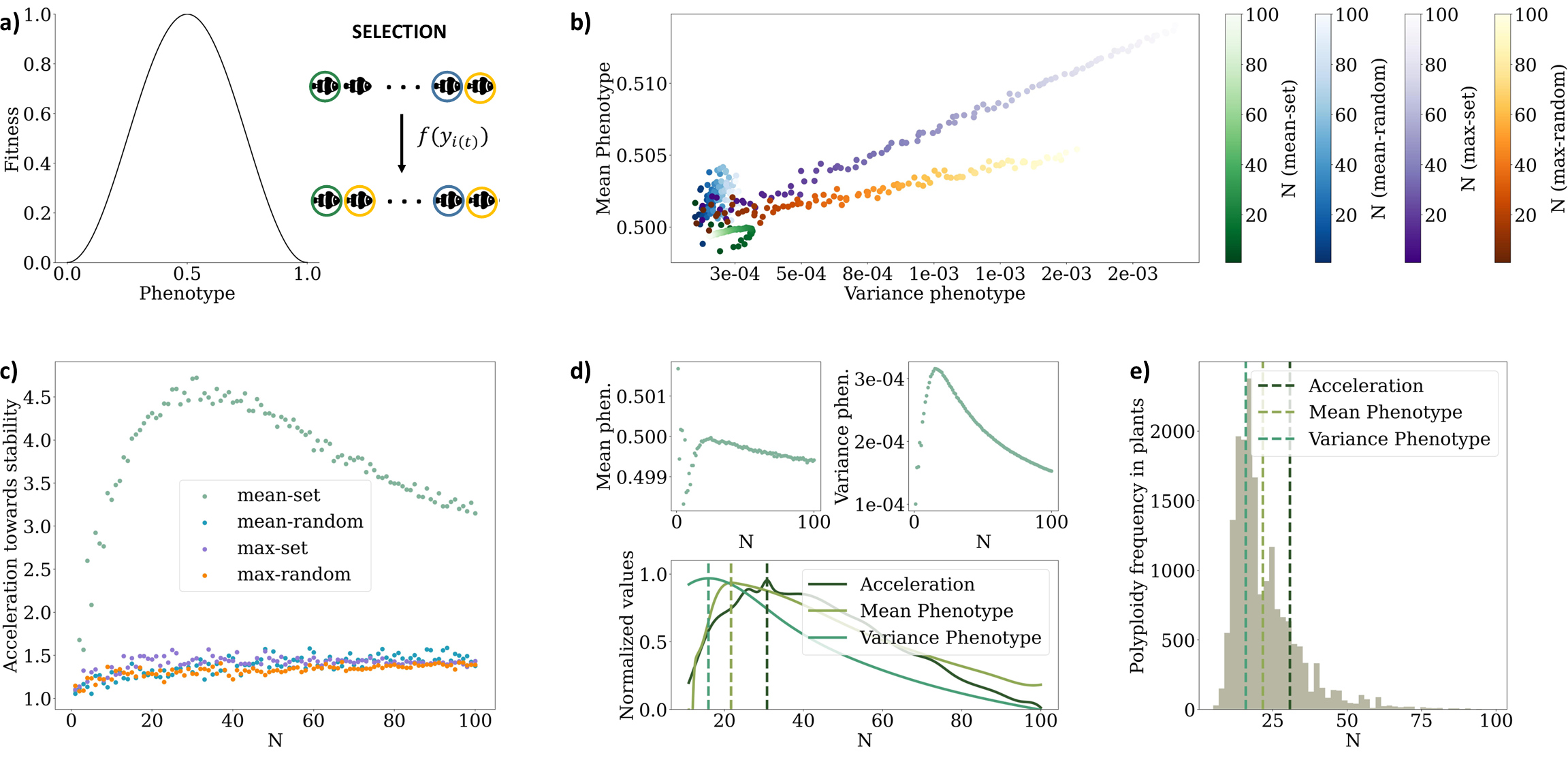}
\caption{\textbf{Evolutionary dynamics under \textit{smooth fitness}.}
\textbf{a)} \textit{Smooth fitness} landscape, where the fitness of the $i$-th individual with phenotype $y_i(t)$ is a sinusoidal function peaked at $y_i=0.5$.
\textbf{b)} Mean and variance of the phenotype at steady state as a function of $N$ for \textit{mean-set} (green), \textit{mean-random} (blue), \textit{max-set} (violet), and \textit{max-random} (orange).
\textbf{c)} Acceleration toward stability $Ac$ (ratio of convergence times with \textit{neutral} vs. \textit{smooth fitness}) as a function of $N$ for each model.
\textbf{d)} On top, mean phenotype (left) and variance (right) at steady state for \textit{mean-set} populations as a function of $N$.
On bottom, normalized and smoothed $Ac$ (dark green), mean phenotype (medium green), and variance of the phenotype (light green) for \textit{mean-set} populations as a function of $N$. Vertical lines mark the $N$ corresponding to the maximum of each quantity.
\textbf{e)} Histogram of the ploidy level of the seed plants from the PloiDB database \cite{halabi2023ploidb}. Vertical lines reproduce the same $N$ values as in panel d).
}
\label{fig2}
\end{figure*}

Having established the baseline behavior in the absence of environmental constraints, we now ask whether the introduction of a specific selective optimum modifies the previous conclusions and, in particular, whether it reveals an optimal ploidy level that would be invisible in the neutral setting.\\
To introduce minimal environmental effects, we consider the \textit{smooth fitness}, shown in Figure \ref{fig2}a.\\
Figure \ref{fig2}b shows the mean and variance of the steady state phenotype as a function of $N$. In both \textit{max} models, the variance increases with $N$, suggesting that nonlinear mappings convert genotypic redundancy into phenotypic diversity, enhancing adaptive potential \cite{miotto2020genome}.
On the other hand, the variance of the \textit{mean} models decreases with $N$ by approximately one order of magnitude. However, the mean phenotype approaches the fitness optimum more closely for both \textit{mean} models at low $N$. Notably, the \textit{mean-set} model shows a pronounced non-monotonic dependence on $N$, with clear inflection points.
This suggests the existence of an optimal ploidy level that balances exploitation accuracy and phenotypic variability.\\
To quantify the susceptibility of adaptation speed to the fitness landscape, we define the acceleration toward stability ($Ac$) as the ratio 
between the number of generations required to reach stability with the \textit{neutral} and in the \textit{smooth fitness}. A value $Ac>1$ indicates 
that the population adapts faster in the latter landscape than in the former, reflecting a more directed evolutionary dynamics. As shown in Figure~\ref{fig2}c, the \textit{max} and \textit{mean-random} models show $Ac$ values that are approximately independent of $N$, 
consistent with their sustained phenotypic exploration across all ploidy levels.
By contrast, the \textit{mean-set} populations show a strongly $N$-dependent $Ac$, with a pronounced peak at $N \approx 30$.
Figure~\ref{fig2}d confirms that the peaks in exploitation (mean phenotype peaking at $N \approx 22$), exploration (variance peaking at $N \approx 16$), and convergence speed (acceleration peaking at $N \approx 30$) all occur at low-to-intermediate $N$ values, forming a coherent picture of an optimal ploidy range for \textit{mean-set} populations.
Notably, these $N$ values are often found in polyploid organisms \cite{simon1977macromolecular,ohbayashi2020evolutionary}.
For example, \textit{Anabaena cylindrica} carries approximately 25 chromosome copies per cell \cite{simon1977macromolecular}, and multiple species of \textit{Cyanobacterium} and \textit{Geminocystis} maintain 16-34 copies \cite{ohbayashi2020evolutionary}. 
Moreover, Figure \ref{fig2}e shows that the distribution of ploidy levels in seed plants (from the PloiDB database \cite{halabi2023ploidb}) is also concentrated in the low-to-intermediate $N$ range, with a  distribution that closely matches the peak region identified in our model. While seed plants are complex multicellular organisms for which many phenotypes are likely better described by more sophisticated models, 
the additive mean-set approximation may be relevant for traits determined by dosage-sensitive genes, which have been found to maintain growth and developmental stability in polyploid plants \cite{shi2015genome}.\\

This suggests that in organisms following the \textit{mean-set} model and living in environments rewarding an intermediate phenotype, different $N$ values can optimize different characteristics of the population evolution such as speed, exploitation, and exploration.\\

These results raise a further question: whether the advantages conferred by polyploidy are amplified when the population must respond not simply to a static optimum, 
but to an abrupt and drastic change in the fitness landscape, which is a scenario directly relevant to the evolutionary conditions under which polyploidization events are most frequently observed in nature \cite{te2012more,hahn2012increased,ramsey2011polyploidy,madlung2013polyploidy}.

\subsection*{Polyploidy enhances phenotypic exploration and facilitates escape from fitness valleys under abrupt environmental change}
\begin{figure*}[ht]
\centering
\includegraphics[width=\linewidth]{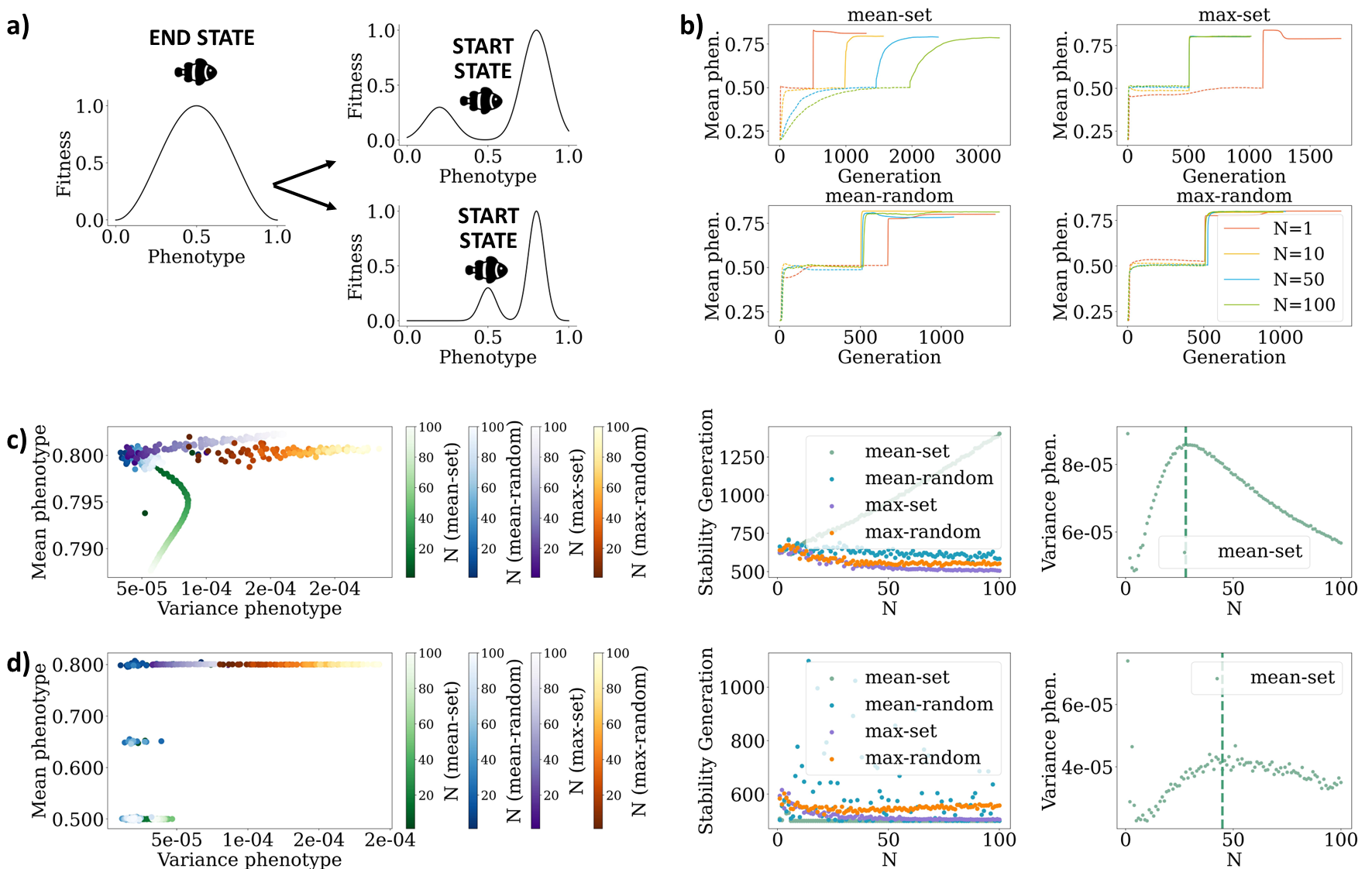}
\caption{\textbf{Adaptability of the four models under abrupt environmental changes.}
\textbf{a)} Schematic representation of the two \textit{rough fitnesses} landscapes used to probe adaptability. With the \textit{valley fitness} (top) the 
initial population, stabilized near $y_i \approx 0.5$ in the smooth landscape, is placed in a fitness minimum between two Gaussian peaks. 
With the \textit{peak fitness} (bottom), the initial population is at a local suboptimal maximum.
\textbf{b)} Mean phenotype during evolution with the \textit{valley fitness} for $N=1,10,50,100$ (red, yellow, blue, and green, respectively) for the four models.
\textbf{c)} On the left, mean and variance of the phenotype at steady state as a function of $N$ for \textit{mean-set} (green), \textit{mean-random} (blue), \textit{max-set} (violet), and \textit{max-random} (orange). At the center, the number of generations required to reach stability as a function of $N$. On the right, variance of the phenotypes for \textit{mean-set} populations as a function of $N$. The populations evolved under the \textit{valley fitness}.
\textbf{d)} Same as in c), for the \textit{peak fitness}.
}
\label{fig3}
\end{figure*}

To investigate how populations respond to an abrupt change in the fitness landscape, we consider a transition from the \textit{smooth fitness} to two distinct \textit{rough fitnesses} (see Methods for more details). This allows us to probe the ability of different inheritance modes and phenotype mappings to escape local optima and explore more complex adaptive landscapes.\\
We consider the two scenarios shown in Figure \ref{fig3}a: (i) with the \textit{valley fitness} the population starts at a fitness minimum for $y_i\sim0.5$ between two peaks (one local maximum for $y_i=0.2$ and a global maximum at $y_i=0.8$), while (ii) with the \textit{peak fitness}, the population starts at a local maximum for $y_i\sim0.5$ and must climb out to the higher global optimum at $y_i = 0.8$.\\

To mimic an environmental shift and assess the adaptability of the different models, after the population has reached a stationary state with the \textit{smooth fitness}, we modify $f(y_i(t))$ and study the role of the different models in driving the adaptation of the population to the new environment.\\
Figure \ref{fig3}b shows representative trajectories of the mean phenotype during evolution for $N = 1, 10, 50, 100$ under the four models with the \textit{valley fitness}.
The \textit{max} models are confirmed to be the most explorative ones (higher variance) in both scenarios, and the fastest to reach stability, as shown in Figure \ref{fig3}c-d. 
Increasing $N$ further enhances exploration in these models, since a larger number of chromosomes increases the probability that at least one copy carries a high-value allele enabling a jump to the new optimum.

The \textit{mean-random} model also converges quickly and reaches the global optimum with the \textit{valley fitness}, but displays lower phenotypic variance and reduced performance with the \textit{peak fitness}. 
The \textit{mean-set} model is confirmed to be the least adaptable model: in the \textit{peak} landscape it fails to reach the global maximum for any value of $N$. 
Nonetheless, it again exhibits a non-monotonic dependence of variance on $N$, with a peak at an intermediate polyploidy level with both the \textit{valley} and \textit{peak fitnesses}. This indicates that even within the least explorative model, there exists an optimal $N$ that maximizes phenotypic exploration, consistent with the pattern observed in the smooth landscape.\\

These results suggest that polyploidy can confer an adaptive advantage during abrupt environmental changes in all models, primarily by expanding 
phenotypic variability. This is consistent with paleontological evidence linking ancient WGD events with major environmental transitions, such as 
the Cretaceous-Paleogene mass extinction \cite{vanneste2014tangled}.

\subsection*{Natural selection on chromosome number favors a balance between exploitation, exploration, and convergence speed}

\begin{figure*}[ht]
\centering
\includegraphics[width=\linewidth]{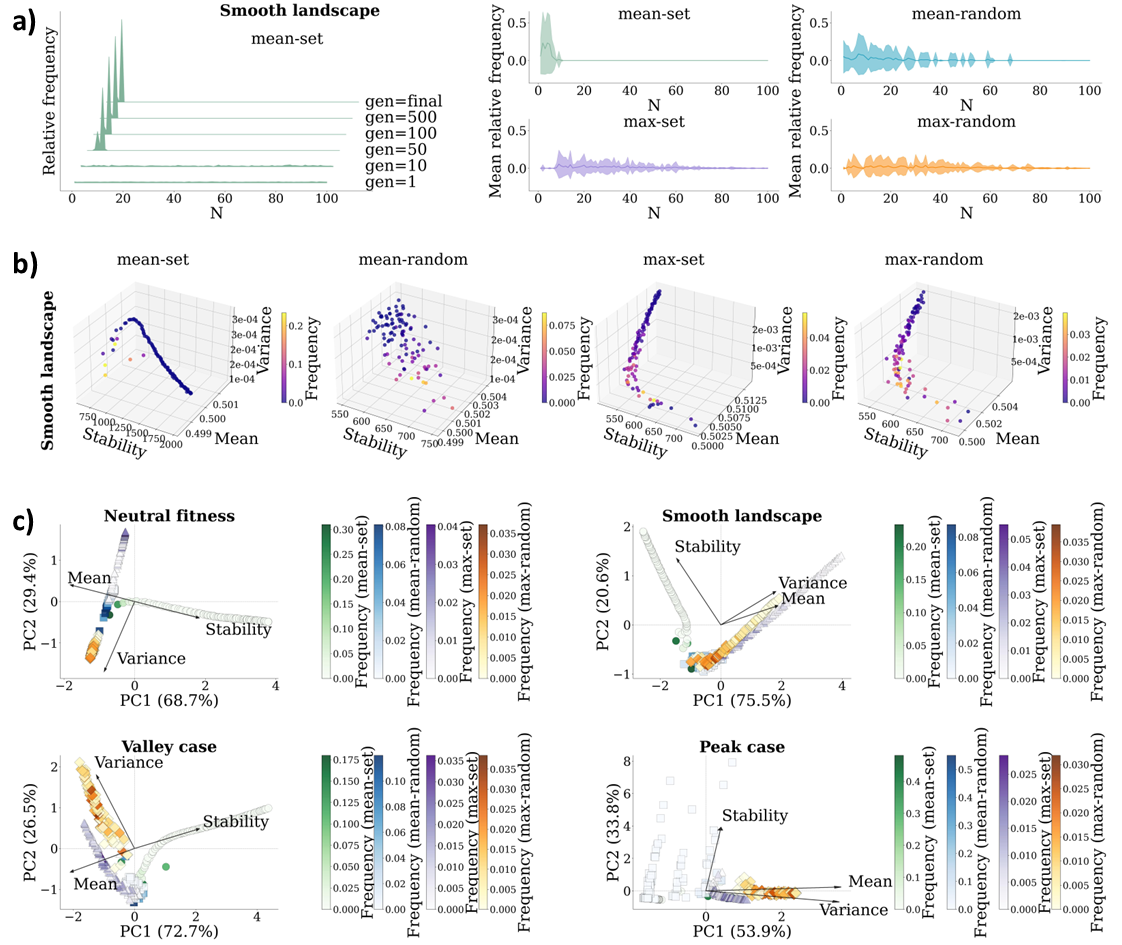}
\caption{\textbf{Identification of evolutionarily successful polyploidy levels in populations with heterogeneous chromosome number.}
\textbf{a)} On the left, distribution of $N$ during evolution in the \textit{smooth fitness} landscape for a \textit{mean-set} population starting from a uniform distribution over $N \in [1,100]$. Each curve corresponds to a different generation.
On the right, distribution of $N$ in the final stable populations for the four models in the \textit{smooth fitness} landscape. Solid lines indicate the mean frequency as a function of $N$, while shaded regions indicate its variance over 100 repetitions. See the Supplementary for the other three fitness landscapes.
\textbf{b)} For each of the four models in the \textit{smooth fitness} landscape, the stability generation, mean phenotype, and variance at steady state are plotted as three-dimensional scatter plots for each fixed-$N$ simulation. Points are colored according to the frequency with which that $N$ appears in the stable mixed-$N$ population. See the Supplementary for the other three models.
\textbf{c)} Projection of the same data onto the first two Principal Components (PC1 and PC2), for each of the four environmental conditions (panels from left to right and top to bottom: \textit{neutral}, \textit{smooth}, \textit{peak}, and \textit{valley fitnesses}). Points are colored by frequency. Black arrows indicate the direction and magnitude of the contribution of each variable to the PCs (see Methods).
}
\label{fig4}
\end{figure*}

Having shown that for each of the four models different ploidy levels confer different evolutionary advantages (either in terms of phenotypic exploitation, phenotypic exploration, and/or convergence speed), we investigate whether there are optimal $N$ values resulting in the general success of a population and what tradeoffs in evolutionary advantages confer this success.\\
With this aim, we modify the model so that each individual in the starting population has equal probability of carrying any $N \in [1,100]$ chromosomes.
Mothers pass their chromosome number $N$ to their offspring unchanged, so that $N$ is an inherited trait subject to indirect selection 
through its effects on phenotype and fitness.\\
Figure \ref{fig4}a shows, as an example, the evolution of the $N$ distribution in a \textit{mean-set} population in the \textit{smooth fitness} landscape. Starting from a uniform distribution over $[1, 100]$, individuals with low $N$ progressively prevail after a few rounds of reproduction.
This is consistent with the results of the fixed-$N$ simulations: while low $N$ values did not confer the best mean phenotype and highest variance, they provided the faster convergence.
For the \textit{max} models, the distribution is less peaked, presumably because of a conflict between exploitation, greater for lower $N$, and exploration, enhanced for higher $N$ (see Figure \ref{fig2}b).\\

To understand which combination of evolutionary features determines the success of each $N$, we perform a Principal Component Analysis (PCA) of the three key quantities (stability generation, mean phenotype, and variance of the phenotype at steady state) computed from the fixed-$N$ simulations (Figure~\ref{fig4}b). For each $N$, we color the corresponding data point according to the frequency with which that $N$ appears in the stable mixed-$N$ population, providing a direct link between evolutionary features and selective success.\\
Not all evolutionary features contribute equally to selective success across environments: to quantify their relative importance, we examine the loading vectors of the PCA (i.e. the contribution of each original variable to the variance explained by the first two PCs, as discussed in more detail in the Methods Section), which reveal which of the three quantities most strongly drives the separation between successful and unsuccessful $N$ values in each condition.\\
Figure \ref{fig4}c shows the loading vectors in the two-dimensional PCs space for stability generation, mean and variance of the phenotype for all models and fitness landscapes.
The direction of each arrow indicates the direction in PC space along which the corresponding variable increases most rapidly, while the length of the arrow is proportional to the magnitude of the loading.
Variables with long arrows in the direction of PC1 are the primary drivers of the dominant source of variation in the dataset; variables with arrows pointing along PC2 capture the secondary source of variation. The angle between two loading vectors approximates the correlation between the corresponding variables: small angles indicate high positive correlation, while angles close to $180^\circ$ indicate high negative correlation.\\
In the \textit{neutral fitness} environment, the most successful $N$ values are primarily those associated with high phenotypic variance, reflecting the dominant role of variability in driving reproductive success when there is no environmental optimum. 
In the \textit{smooth fitness} landscape, variance and mean of the phenotype are determinant.
In both the \textit{valley} and \textit{peak fitnesses}, phenotypic variance again becomes the dominant driver, but the convergence speed also contributes, particularly in the \textit{mean-set} model where the stability generation shows a pronounced $N$-dependence.\\

This suggests that the selective advantage of a given polyploidy level is context-dependent, consistent with empirical observations that optimal ploidy levels vary substantially across species and ecological settings, as reviewed by Otto, S. P. and Whitton, J. \cite{otto2000polyploid} and Comai, L. \cite{comai2005advantages}.

\section*{Conclusions}

Different inheritance patterns and genotype-phenotype mapping can significantly affect genetic variance and ultimately evolutionary dynamics.
Understanding the role of polyploidy in these different conditions and its interplay with natural selection has many practical applications. For example, cancer is increasingly seen as an evolutionary system in which cells are subject to both a strong natural selection (from the immune system and anticancer therapies) and high entropic forces (fast mutation rates) \cite{laplane2024evolutionary}, and polyploidy has been ascribed an increasing role in tumor evolution and drug resistance \cite{van2017evolutionary}.\\
We introduce a minimal one-dimensional evolutionary model to investigate how polyploidy, inheritance mode, and genotype-phenotype mapping jointly determine the optimization and adaptability of populations under varying selective pressures.\\
We find that (i) the combination of stochastic inheritance and a 
nonlinear (maximum-based) phenotype mapping consistently maximizes both phenotypic exploitation and landscape exploration, providing the broadest adaptive potential across all fitness landscapes considered.
(ii) In the \textit{mean-set} model, which captures additive gene expression under structured inheritance, a clear optimal ploidy level emerges in the range $N \approx 15$--$30$, at which populations simultaneously maximize exploitation, exploration, and convergence speed. This prediction is quantitatively consistent with the ploidy distributions observed in polyploid bacteria, cyanobacteria, and seed plants \cite{simon1977macromolecular,ohbayashi2020evolutionary,halabi2023ploidb}, suggesting that our minimal model captures an ecologically relevant selection pressure. 
(iii) When the fitness landscape undergoes an abrupt change,  polyploidy enhances phenotypic exploration and facilitates escape from suboptimal fitness states in all four models, supporting the long-standing hypothesis that WGD events are preferentially associated with environmental upheaval \cite{te2012more,hahn2012increased,ramsey2011polyploidy,
madlung2013polyploidy}.
(iv) When chromosome number is itself subject to indirect selection, the evolutionarily successful ploidy levels reflect a context-dependent balance between exploitation, exploration, and convergence speed, rather than any single evolutionary objective.\\

Several limitations of our model should be acknowledged. The one-dimensional, single-locus architecture cannot correctly reproduce the complexity of real genomes, including recombination, epistasis, and linkage. The two inheritance modes considered are idealizations of a continuous spectrum from disomic to polysomic segregation: model-independent methods that quantify hereditary structure directly from lineage tree data \cite{allegrezza2026inheritance} could provide a principled way to position real biological systems along this spectrum and to test the predictions of our framework against experimentally reconstructed lineages.
Our genotype–phenotype mapping is also simplified, as real systems may exhibit a wide spectrum of relationships between fully additive (linear) and maximally non-additive (extreme-value–like) contributions of gene copies. Moreover, real fitness landscapes are typically multidimensional and time-varying.
Future work should investigate how these factors modify the predictions derived here, and how the evolutionary dynamics of $N$ interact with mutation rate evolution and genome stability.
Coupling our framework with experimentally derived lineage topologies and replication kinetics, such as those characterized in stromal cell colonies \cite{allegrezza2026lineage}, could help bridge the gap between our minimal protocol and the regulated growth dynamics observed in real proliferating cell populations.

\section*{Materials and Methods}

\subsection*{One-dimensional evolutionary model}
Each simulation was initialized with the following parameters:
\begin{itemize}
    \item Population size $M=10,000$ individuals.
    \item Initial gene copy values $x_{ij}(t=0)=0.2$ for all individuals
    \item Mutation rate $\lambda=0.00032$ per gene copy per generation. The number of mutated genes per individual per generation is drawn from a Poisson distribution with mean $\lambda N$.
     \item Mutated gene values are reassigned by sampling from a Beta distribution $\mathrm{Beta}(1, \beta)$ with $\beta = 1/\bar{x} - 1$, where $\bar{x} = 0.5$ is the mean 
    of the mutation distribution. This ensures that mutations explore the full range $[0,1]$ with a bias toward intermediate values.
\end{itemize}

The \textit{rough fitness} landscapes were defined as follows:
\begin{itemize}
    \item The \textit{valley fitness} as $f(y_i)=A_v\mathcal{N}(0.2,0.09)+\mathcal{N}(0.8,0.09)$, with $A_v=0.3$.
    \item The \textit{peak fitness} as $f(y_i)=A_p\mathcal{N}(0.5,0.05)+\mathcal{N}(0.8,0.05)$, with $A_p=0.3$.
\end{itemize}

\subsection*{Principal Component Analysis and loading vectors}
To identify the combination of evolutionary features that determines the selective success of each ploidy level $N$, we performed a Principal Component Analysis (PCA) on the three quantities characterizing each fixed-$N$ simulation: the stability generation, the mean phenotype at steady state, and the variance of the phenotype at steady state. For each environmental condition and each model, these three quantities were collected across all $N \in [1, 100]$, standardized to zero mean and unit variance (z-score normalization), and then subjected to a PCA.
PCA is a multivariate statistical technique that transforms the original basis vectors describing the dataset into an orthogonal basis formed by the eigenvectors (also known as PCs) of the sample covariance matrix $\hat{C}$. Each PC is associated to an eigenvalue, indicating the amount of variance it captures from the data. The PCs can be ranked in descending order based on their eigenvalues; by selecting the first $d$ PCs, the dimensionality of the dataset can be reduced while preserving its dominant structure.\\
Each PC is given by a linear combination of the original variables. The contribution of each original variable to a PC is called loading, and we define the loading vector of variable $k$ as $(v_{k1},..., v_{kd})$, where $v_{k\ell}$ is the $k$-th component of the $\ell$-th eigenvector of $\hat{C}$. 
Analyzing the loading vector can shed light on the interplay between the features of the original dataset in determining a particular condition of its components \cite{grassmann2026exploring}.
In our representation, the arrows are also scaled by a factor proportional to the range of the projected scores, so that their length is visually comparable to the spread of the data points in the plot.

\section*{Data Availability}
The data that support the findings of this study are available from the corresponding author upon request.

\section*{Declaration of interests}
The authors declare no competing interests.

\section*{Acknowledgements}
MM acknowledges the contribution of the Italian Ministero dell’Università e della Ricerca, Decreto Ministeriale No. 1236 of August 1, 2023—FIS 2 CALL. Project FIS2023-02957 (CUP B53C24009530001), ’Fathoming oUt the role of partitioninG noIse in cancer epithelial-mesenchymal Transitions (FUGIT).

\bibliographystyle{unsrt}
\bibliography{mybibfile}

\end{document}


\section{Supplementary Information}

\begin{figure*}[ht]
\centering
\includegraphics[width=\linewidth]{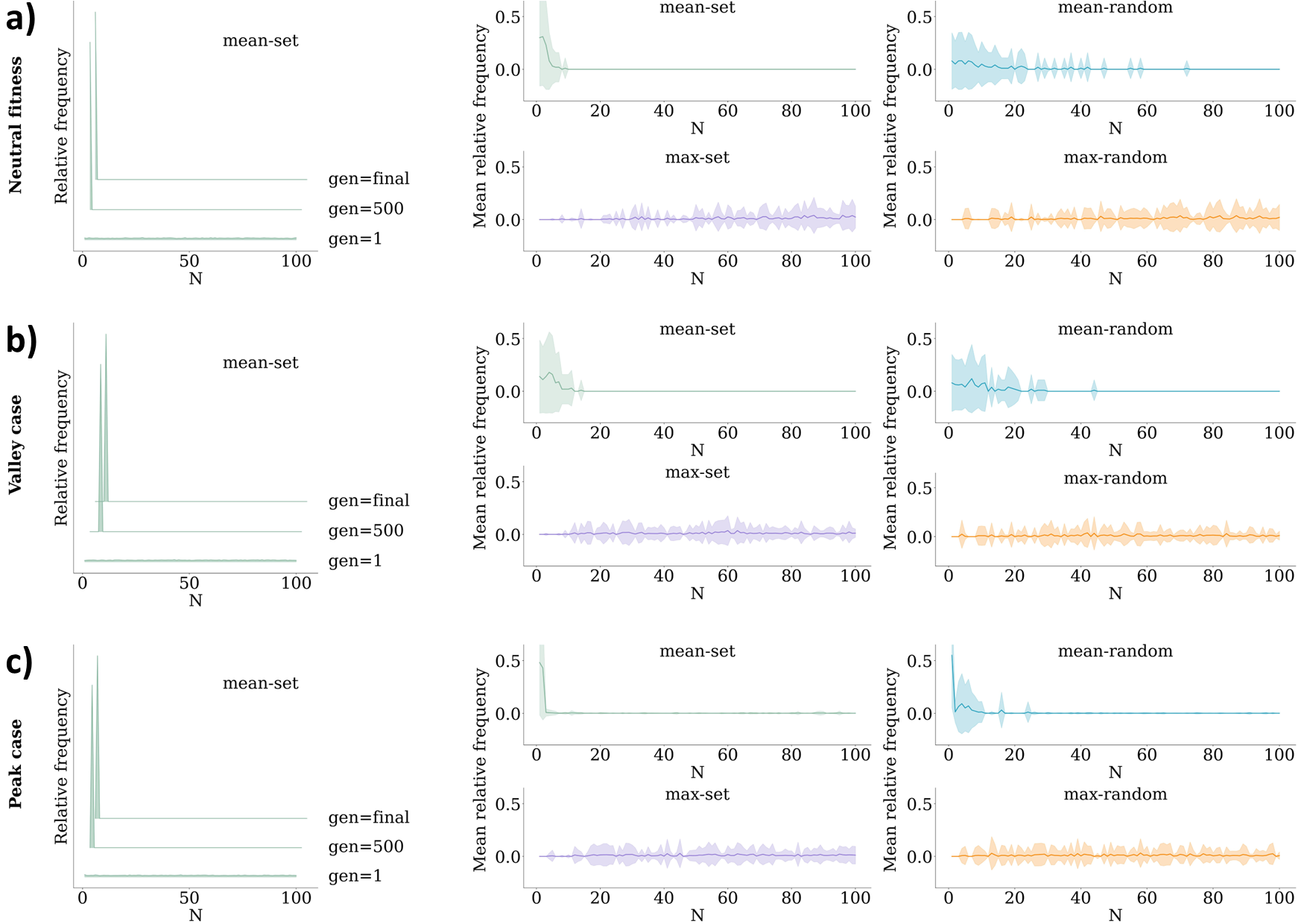}
\caption{\textbf{$N$ distribution at steady state in the mixed-$N$ simulations.}
\textbf{a)} On the left, distribution of $N$ during evolution in the neutral fitness landscape for a \textit{mean-set} population starting from a uniform distribution over $N \in [1,100]$. Each curve corresponds to a different generation.
On the right, distribution of $N$ in the final stable populations for the four models in the neutral fitness landscape. Solid lines indicate the mean frequency as a function of $N$, while shaded regions indicate its variance over 100 repetitions. 
\textbf{b)} Same as in a) for the \textit{valley} case.
\textbf{c)} Same as in a) for the \textit{peak} case.
}
\label{S1}
\end{figure*}

\begin{figure*}[ht]
\centering
\includegraphics[width=\linewidth]{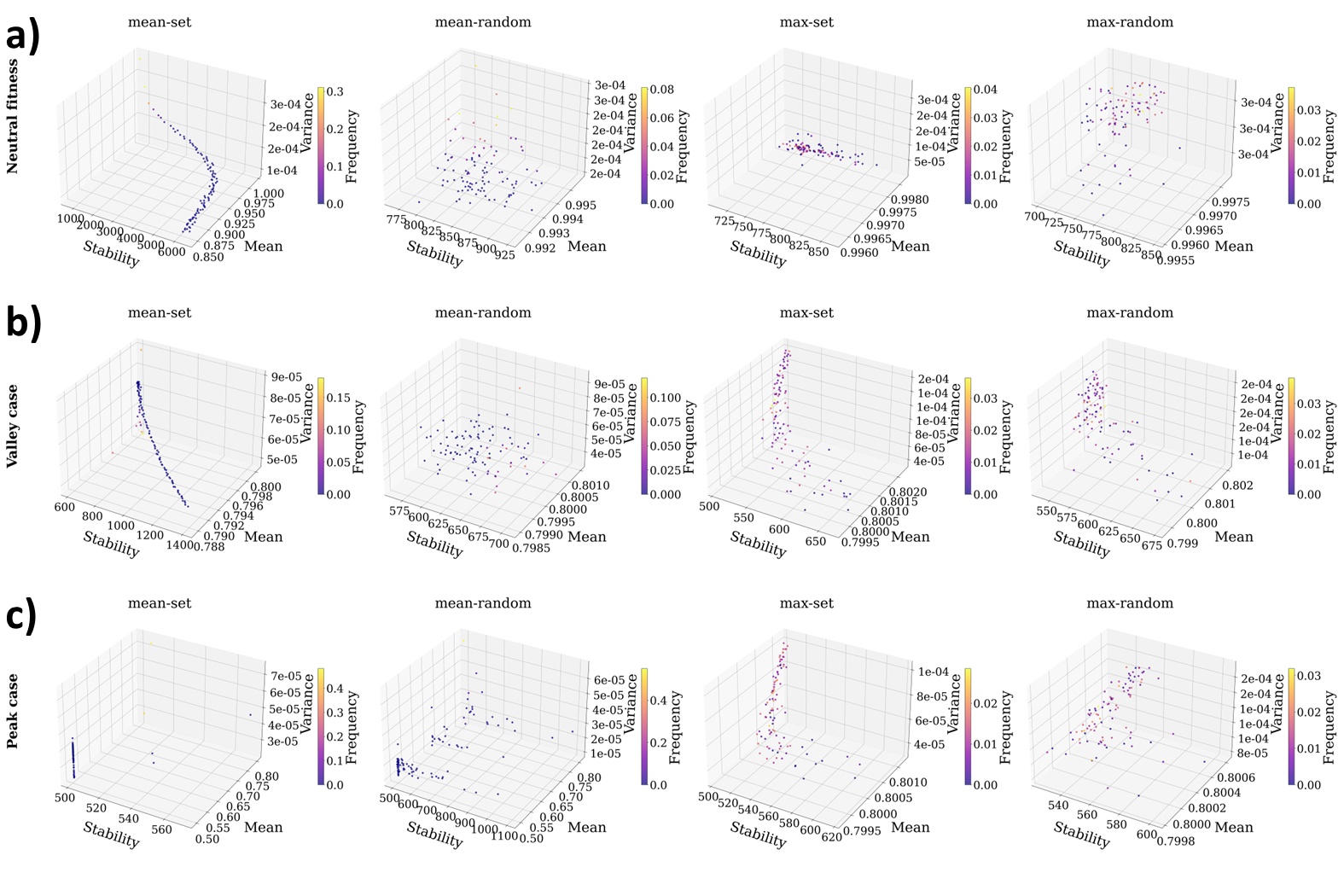}
\caption{\textbf{Evolutionary features of fixed-$N$ populations and success of the $N$ ploidy levels.}
\textbf{a)} For each of the four models (from left to right, \textit{mean-set}, \textit{mean-random}, \textit{max-set}, and \textit{max-random}) in the neutral fitness landscape, the stability generation, mean phenotype, and variance at steady state are plotted as three-dimensional scatter plots for each fixed-$N$ simulation. Points are colored according to the frequency with which that $N$ appears in the stable mixed-$N$ population. 
\textbf{b)} Same as in a) for the \textit{valley} case.
\textbf{c)} Same as in a) for the \textit{peak} case.
}
\label{S2}
\end{figure*}